\documentclass[twocolumn,preprintnumbers,amsmath,amssymb,superscriptaddress]{revtex4}
\usepackage{array}
\usepackage{booktabs}
\usepackage{tabu}
\usepackage{dcolumn}
\usepackage{amsmath}
\usepackage{amsfonts}
\usepackage{amssymb}
\usepackage{graphicx}
\usepackage[colorlinks,linkcolor=red,citecolor=blue]{hyperref}
\usepackage{subfigure}
\usepackage{graphicx}
\usepackage{dcolumn}
\usepackage{bm}
\usepackage{dsfont}
\newtheorem{thm}{Theorem}

\newcommand{\commentold}[1]{}
\DeclareMathSymbol{:}{\mathpunct}{operators}{"3A}

\def\be{\begin{equation}}
\def\ee{\end{equation}}
\def\bea{\begin{eqnarray}}
\def\eea{\end{eqnarray}}
\def\f{\frac}
\def\n{\nonumber}
\def\l{\label}

\begin{document}
\title{No Entropy Production in Quantum Thermodynamics}
\author{B. Ahmadi}
\email{b.ahmadi19@gmail.com}
\address{Department of Physics, University of Kurdistan, P.O.Box 66177-15175, Sanandaj, Iran}
\address{International Centre for Theory of Quantum Technologies, University of Gdansk, Wita Stwosza 63, 80-308 Gdansk, Poland}
\author{S. Salimi}
\email{shsalimi@uok.ac.ir}
\address{Department of Physics, University of Kurdistan, P.O.Box 66177-15175, Sanandaj, Iran}
\author{ A. S. Khorashad}
\address{Department of Physics, University of Kurdistan, P.O.Box 66177-15175, Sanandaj, Iran}
\date{\today}

\def\br{\biggr}
\def\bl{\biggl}
\def\Br{\Biggr}
\def\Bl{\Biggl}
\def\be\begin{equation}
\def\ee{\end{equation}}
\def\bea{\begin{eqnarray}}
\def\eea{\end{eqnarray}}
\def\f{\frac}
\def\n{\nonumber}
\def\l{\label}
\begin{abstract}
In this work we will show that there exists a fundamental difference between microscopic quantum thermodynamics and macroscopic classical thermodynamics. It will be proved that the entropy production in quantum thermodynamics always vanishes for both closed and open quantum thermodynamic systems. This novel and very surprising result is derived based on the genuine reasoning Clausius used to establish the science of thermodynamics in the first place. This result will interestingly lead to define the generalized temperature for any non-equilibrium quantum system.
\end{abstract}

\keywords{Suggested keywords}
\maketitle
\begin{figure}[h]
\centering
\includegraphics[width=4cm]{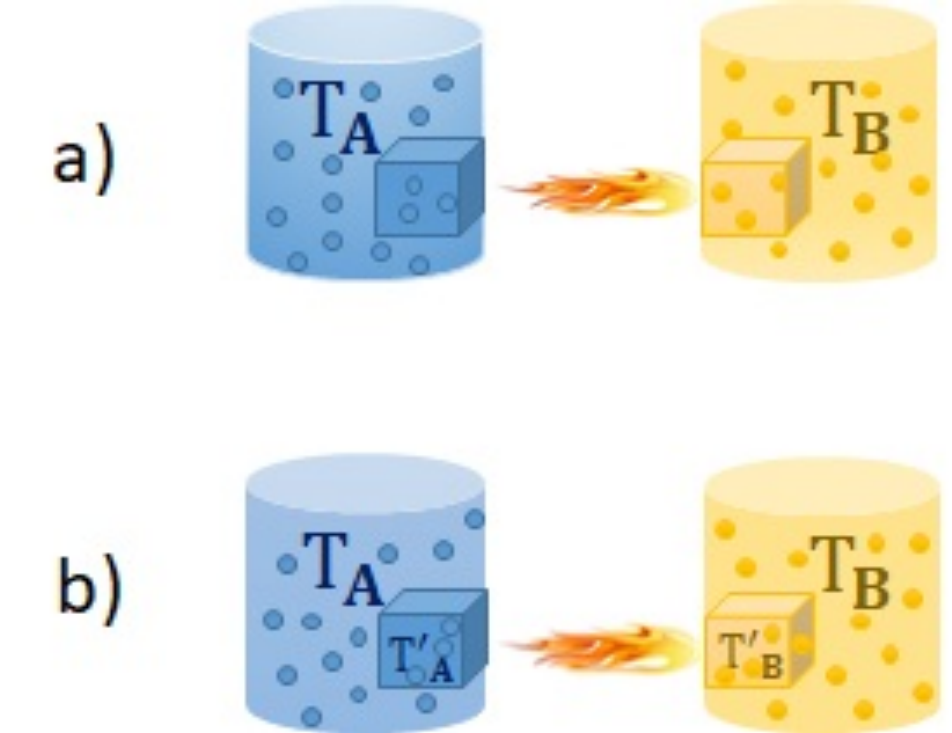}
\caption{(Color online) Two subsystems $A$ and $B$ with temperatures $T_A<T_B$ are interacting. a) At time $t=0$ where the subsystems begins to exchange heat. b) After time $\tau$ where the two subsystems are in non-equilibrium states.}
\label{fig:Fig1.eps}
\end{figure}
\begin{figure}[h]
\centering
\includegraphics[width=3cm]{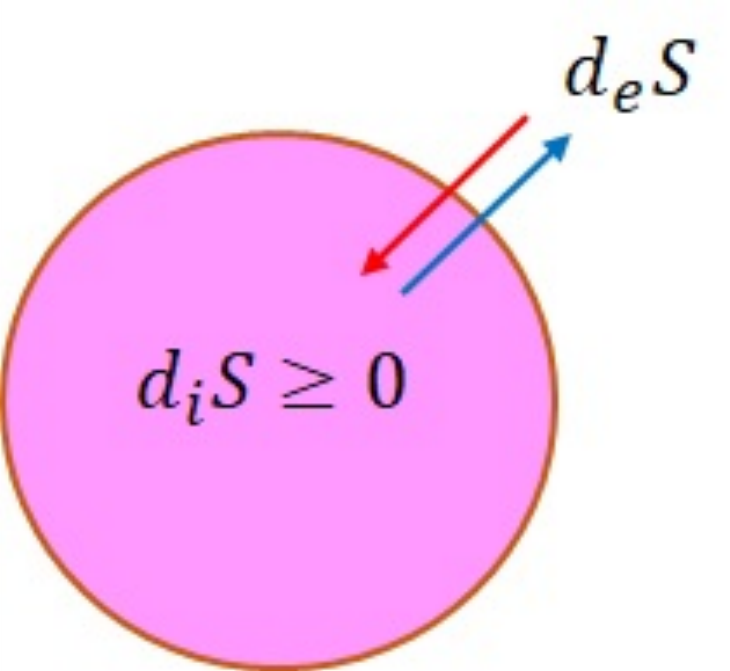}
\caption{(Color online) The total entropy change of any thermodynamic system is divided into two parts: $dS=d_eS+d_iS$. $d_eS$ comes from the interaction with the environment and can be either positive or negative. $d_iS$ is produced in the interior of the system and is always positive for any thermodynamic system.}
\label{fig:Fig5}
\end{figure}
\begin{figure}[h]
\centering
\includegraphics[width=7cm]{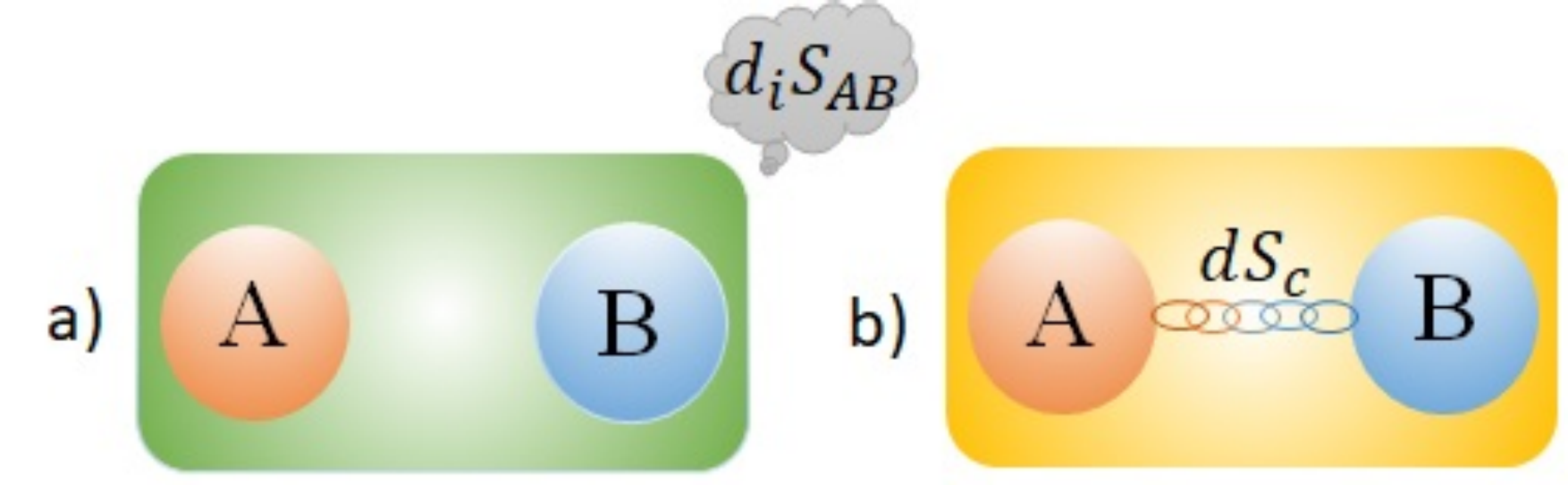}
\caption{(Color online) A closed system $AB$ made of two interacting subsystems $A$ and $B$. a) Two classical subsystems interacting with each other and as a result information may leave the total system $AB$, i.e., $dS_{AB}=d_iS_{AB}\geq0$. b) Two quantum subsystems interacting with each other and due to correlations established between them no information can leave the total system $AB$, i.e., $dS_{AB}=0$.}
\label{fig:Fig2.eps}
\end{figure}

\newpage
\textit{Introduction}. Entropy production is the entropy which is produced in the interior of a system. It is produced even if the \textit{classical} system is isolated from the surroundings. The Clausius statement of the second law of classical thermodynamics asserts that the entropy production of a system is either zero or positive \cite{Kondepudi,Groot,Bellac,Yunus}. It is zero for systems in equilibrium and positive for non-equilibrium systems. A process is thus thermodynamically reversible if and only if the entropy production is zero \cite{Kondepudi,Groot,Bellac,Yunus} and this means that a reversible process is a process that can be reversed without leaving any trace on the surroundings, i.e., leaving the entropy of the surroundings unchanged (no matter how the process is reversed) \cite{Kondepudi,Groot,Bellac,Yunus}. Since in a \textit{macroscopic classical} thermodynamic system entropy may be produced in the interior of the system therefore it is expected that the same is true for a \textit{microscopic quantum} thermodynamic system. But wrong presumptions may lead to misconceptions. Hence before computing the entropy production of a quantum thermodynamic system we first need to make sure that the mechanics of a quantum system is capable of creating such entropy inside the system.
\newline
In this work we will show that, as expected, fundamental differences appear in quantum thermodynamics. It will be proved that the entropy production always vanishes for quantum systems as a result of the laws of quantum mechanics. We will see that the information which is supposed to get lost is in fact stored in the quantum correlations established between subsystems. It also results in defining the generalized temperature for non-equilibrium quantum systems.
\newline
\textit{Entropy production in classical thermodynamics}. Consider two classical systems $A$ and $B$ which are in equilibrium with temperatures $T_A<T_B$ at time $t=0$. Part of the system $B$ begins to exchange heat quasi-statically with part of the system $A$ (as shown in Fig. (\ref{fig:Fig1.eps})). As time passes the two systems become non-equilibrium and this is because part of system $A$ still has temperature $T_A$ and its another part has temperature $T'_A$ with $T_A\neq T'_A$ (see illustration in Fig. (\ref{fig:Fig1.eps}b)). The same happens to system $B$. Now the change in the entropy of each system, from $t=0$ to $t=\tau$, due to the interaction with another system is obtained as \cite{Kondepudi,Groot,Bellac,Yunus}
\begin{equation}\label{1}
\Delta_eS_{A(B)}=\int_{0}^{\tau}\dfrac{dQ_{A(B)}}{T'_{A(B)}(t)},
\end{equation}
where $dQ$ is the exchanged heat and $T'$ the temperature of that part of the system which is in contact with another system and the letter $e$ stands for the word "exterior", i.e., $\Delta_eS$ occurs in the exterior of the systems. But this is not the whole change in the entropy of the systems. The two systems are in non-equilibrium states due to the temperature gradient inside themselves and from \textit{experimental observations} we know that any non-equilibrium macroscopic classical system \textit{tends} to reach a new equilibrium. And since the equilibrium state has a higher entropy therefore we can say that any non-equilibrium system tends to increase its entropy. It should be noticed that this \textit{tendency} has nothing to do with the interaction of the system of interest with another system therefore it also holds for any closed non-equilibrium macroscopic classical system. What another system (environment) does on the system of interest is that it causes the system of interest to become non-equilibrium which is not enough for entropy to be produced in the interior of the system. There should also exist a tendency inside the system to go from this non-equilibrium state toward an equilibrium state with a higher entropy. This is the most crucial point in thermodynamics and we will go back to it in the quantum case. Even if we cut off the interaction between the two systems, at any time, this tendency still remains in both systems and after some time they will reach a new equilibrium. In other words entropy is produced in the "interior" of each system, i.e., $\Delta_iS_{A(B)}\neq 0$ and it is called \textit{entropy production} which is always positive for any classical thermodynamic system (see Fig. (\ref{fig:Fig5})) \cite{Kondepudi,Groot,Bellac,Yunus}. The letter $i$ stands for the word "interior". Hence the total change in the entropy of the system (for instance $A$) reads
\begin{equation}\label{2a}
\Delta S_A=\Delta_eS_A+\Delta_iS_A.
\end{equation}
The difference between $\Delta_eS$ and $\Delta_iS$ is that the former is caused by the interaction between the two systems on the border while the latter is caused by the interactions in the interior of each system. In other words $\Delta_eS$ is caused by the heat flow in the exterior of the system and $\Delta_iS$ is caused by the heat flow in the interior of the system.
\newline
Therefore for entropy production of a system to be nonzero: "the system must first be in a non-equilibrium state and then the interactions inside the system should be such that a tendency to produce entropy is created and as a result the entropy of the system increases". Since interactions inside the system are such that an isolated macroscopic non-equilibrium classical system always tends to evolve \textit{spontaneously} towards a single stationary state (equilibrium state) 
the entropy of the system changes (increases). In classical thermodynamics this is called the Clausius' statement of the second law of thermodynamics which asserts that for \textit{any} (closed or open) system we have \cite{Kondepudi,Groot,Bellac,Yunus}
\begin{equation}\label{2}
d_iS\geq 0.
\end{equation}
The Clausius' statement of the Second Law can be interpreted that the interactions in the "interior" of a classical system are such that the inequality (\ref{2}) is always satisfied on a macroscopic level. The interactions inside the system could have been such that the entropy production was always zero or even negative. If the entropy production was always zero then the interactions inside the system were such that an isolated system which is in a non-equilibrium state would never necessarily reach equilibrium because as we know the entropy of an equilibrium state is always larger than that of a non-equilibrium state. And if the entropy production was always negative the interactions inside the system were such that the free expansion of a classical gas would never occur. In the following we will show that in quantum thermodynamics the microscopic interactions in the interior of the quantum system are such that no entropy can be produced inside the quantum system.
\newline
\textit{Entropy production in quantum thermodynamics}. In 1932 von Neumann designed a thought experiment to determine the entropy of the density operator $\rho$ of a quantum system \cite{Neumann}. He found out that the entropy of a quantum state $\rho$ reads (see Appendix \ref{AppendixA})
\begin{equation}\label{3}
S(\rho)=:-tr\{\rho\ln\rho\}.
\end{equation}
Now as in the classical case we need to check how much entropy is produced by the interaction with the outside and how much is produced in the interior of the system due to the interactions inside the system. Before doing this job, it should be mentioned that in Ref. \cite{Deffner} the authors have made a mistake in the derivation of the entropy production for non-equilibrium closed quantum systems. In deriving Eq. (\textcolor{red}{12}) from Eq. (\textcolor{red}{7}), in the mentioned paper, they have dropped the second term on the RHS of Eq. (\textcolor{red}{7}) (by mistake!) for no specific reason therefore they have ended up with positive entropy production and this Eq. (\textcolor{red}{12}) has been referred to, for 9 years, by many authors \cite{Binder,Deffner2,Plastina,Santos,Francica,Mancino,Souza,Gherardini,Niedenzu,Francesco,Deffner3,Micadei,Wu}! While as we show below the entropy production of a closed quantum system is \textit{always} zero. Since the evolution of a closed quantum system is unitary then the change in its total entropy is zero and since no heat is exchanged with the outside therefore using Eq. (\ref{2a}) the entropy production of a closed quantum system must be zero. And this will consequently lead to the fact that the entropy production is also zero for open quantum systems because, as we mentioned earlier, the entropy production originates in the \textit{tendency} of the system to increase its entropy from the inside and this tendency has nothing to do with the outside. A closed quantum system which is in a non-equilibrium state does \textit{not tend} to approach an equilibrium state after a long time. The dynamics of a closed quantum system is unitary thus it cannot evolve towards a single stationary state, it actually evolves towards a state from which the information about the initial configuration can be extracted \cite{Breuer,Ahmadi}. This simply suggests that \textit{unlike} classical thermodynamics the entropy production in quantum thermodynamics for closed (and thus open quantum systems) is zero. We will elaborate this issue in the following from every point of view. Consider two quantum systems $A$ and $B$ which are in equilibrium with temperatures $T_A<T_B$ at time $t=0$. The two system begins to interact and thus after some time $t'$ they become non-equilibrium with states $\rho_A(t')$ and $\rho_B(t')$, respectively. Due to the exchanged heat between the systems $\Delta_eS$ may be nonzero for both systems. As for the entropy production, it is expected that some entropy will be produced in the interior of the systems since the systems are in non-equilibrium states. But as we mentioned, in the previous section, being in a non-equilibrium state is not enough for a system to produce entropy on the inside because microscopic interactions inside the system play the major role. As a matter of fact microscopic interactions inside a quantum system are such that the evolution of the state of the system caused by these interactions is unitary thus no entropy may be produced in the interior of the system. And if we cut off the interaction between the two quantum systems at any time both two systems never tend to reach new equilibrium states. These facts simply suggest that no entropy may be produced in the interior of any quantum system. In order to elucidate the argument above we resort to the following discussion. Consider a quantum system $A$ which may generally consists of more than one subsystem, i.e., we may have
\begin{equation}\label{4}
\rho_A=\rho_{12...},
\end{equation}
with Hamiltonian
\begin{equation}\label{5}
H_A(t)=H_1(t)+H_2(t)+...+H_{12}(t)+...
\end{equation}
Any local-in-time master equation, for a quantum system $A$ having a $d$-dimensional Hilbert space, can be written in the form \cite{Hall}
\begin{equation}\label{6}
\dot\rho_A=-\dfrac{i}{\hbar}[H_A(t),\rho_A]+\mathcal{D}(\rho_A).
\end{equation}
where
\begin{equation}\label{7}
\mathcal{D}(\rho_A)=\sum_{k=1}^{d^2-1}\gamma_k(t)[L_k(t)\rho_AL_k^\dag(t)-\dfrac{1}{2}\{L_k^\dag(t)L_k(t),\rho_A\}],
\end{equation}
is the dissipative part of the dynamics in which the $L_k(t)$ form an orthogonal basis set of traceless operators, i.e.,
\begin{equation} \label{8}
tr\{L_k(t)\}=0,\ tr\{L_j^\dag(t)\, L_k(t)\}=\delta_{jk}.
\end{equation}
The first term on the RHS of Eq. (\ref{6}) is the unitary change in the state of system $A$ caused by $H_A(t)$ and $\mathcal{D}(\rho_A)$ the non-unitary change caused by the interaction with the outside. Therefore using Eqs. (\ref{3}) and (\ref{6}) the rate of the change in the entropy of system $A$ becomes
\begin{eqnarray}\label{9}\nonumber
\dot{S}(\rho_A)&=&-tr\{\dot{\rho_A}\ln\rho_A\}\\ \nonumber
&=&tr\{\dfrac{i}{\hbar}[H_A(t),\rho_A]\ln\rho_A\}-tr\{\mathcal{D}(\rho_A)\ln\rho_A\}.\\
\end{eqnarray}
As can be seen the first term on the RHS of Eq. (\ref{9}) is the entropy rate produced inside system $A$ and it is always zero
\begin{equation} \label{9a}
\dfrac{d_iS_A}{dt}=tr\{\dfrac{i}{\hbar}[H_A(t),\rho_A]\ln\rho_A\}=0.
\end{equation}
And the second term is the entropy rate caused by the interaction in the exterior of the system
\begin{equation} \label{9b}
\dfrac{d_eS_A}{dt}=-tr\{\mathcal{D}(\rho_A)\ln\rho_A\}.
\end{equation}
Hence as can be seen unlike the classical thermodynamics in quantum thermodynamics whether the quantum system is in an equilibrium state or not no entropy can be produced in the interior of the system. Now we are in a position to state the following theorem.
\begin{thm}\label{Theorem1}
The entropy production of any quantum thermodynamic system is always zero, i.e.,
\begin{equation}\label{10}
d_iS=0.
\end{equation}
\end{thm}
Theorem \ref{Theorem1} implies the fact that heat flows in the interior of quantum systems do not lead to any increase in the entropy of the system. It must be mentioned that in extending Eq. (\ref{1}) from classical thermodynamics into quantum thermodynamics an obvious mistake has been made in the literature by the authors! When two classical systems $A$ and $B$ with temperatures $T_A$ and $T_B$ quasi-statically exchange heat the change in their entropies are $d_eS_A=dQ_A/T_A$ and $d_eS_B=dQ_B/T_B$, respectively. But if one of the systems, for example system $A$, becomes non-equilibrium since there exists no $T_A$ associated with this system anymore we cannot use the formula $d_eS_A=dQ_A/T_A$. We also cannot solve this problem by replacing $T_A$ with $T_B$, i.e., $d_eS_A=dQ_A/T_B$ to compute the entropy change of system $A$. Instead we should find a hypothetical equilibrium path which produces the same change in the entropy of system $A$ \cite{Kondepudi,Groot,Bellac,Yunus}. This is true because entropy is a function of state and therefore for the purpose of calculations, we can take another path from the initial state to the final state. While H. Spohn \cite{Spohn}, for the first time in deriving entropy production in the weak coupling limit, and then all other authors \cite{Alicki,Breuer,Deffner} have used the formula $d_eS_A=dQ_A/T_B$ for a non-equilibrium quantum system $A$ interacting with a heat bath $B$ at temperature $T_B$. Authors have used the formula $d_eS_A=dQ_A/T_B$ and Alicki's definition of heat \cite{Alicki}, $dQ_A\equiv tr\{d\rho_AH_A\}$, then using Eq. (\ref{3}) they have defined the entropy production of a non-equilibrium quantum system interacting with a heat bath $B$ in equilibrium at temperature $T_B$ as
\begin{equation}\label{11}
d_iS_A=: \ dS(\rho_A)-\dfrac{dQ_A}{T_B}.
\end{equation}
According to Eq. (\ref{11}) since $dS(\rho_A)$ can be different from $dQ_A/T_B$ thus $d_iS_A$ can be nonzero. In Ref. \cite{Esposito} the entropy production has been defined for a quantum system $A$ interacting strongly with a reservoir initially in thermal equilibrium with temperature $\beta=1/T$. They have assumed that $\Delta_eS_A=-\beta\Delta Q_r$ where $\Delta Q_r=tr\{\rho_r(t)H_r-\rho_r(0)H_r\}$ is the heat transferred to the reservoir through the interaction and $\rho_r(0)=\rho_r^\beta$ while $\rho_r(t)$ is a non-equilibrium state. And then using Eq. (\ref{2a}) they have defined the entropy production of the system $A$. The problem with their work is that the assumption $\Delta_eS_A=-\beta\Delta Q_r$ is only true if the system and the reservoir are in thermal equilibrium \textit{during the whole process} \cite{Kondepudi,Groot,Bellac,Yunus} which is not the case in Ref. \cite{Esposito}, and if during the whole process the system is in a thermal equilibrium state no entropy can be produced inside the system.
\newline
Let us now discuss the role of quantum correlations in quantum thermodynamics. In classical thermodynamics, for a closed combined system $AB$ we always have \cite{Kondepudi,Groot,Bellac,Yunus}
\begin{equation}\label{10b}
dS_{AB}=dS_A+dS_B=d_iS_{AB}\geq0,
\end{equation}
i.e., some information leave the total classical system $AB$ due to the irreversibility between $A$ and $B$ and this irreversibility may come from the gradient of the temperature inside the total system $AB$, i.e., $T_A\neq T_B$ (see illustration in Fig. (\ref{fig:Fig2.eps})). Note that $d_iS_{AB}$ in Eq. (\ref{10b}) can still be nonzero even when the two classical subsystems $A$ and $B$ with temperatures $T_A$ and $T_B$ are in equilibrium states when they are exchanging heat, i.e., the change in their entropies are $dS_A=dQ_A/T_A$ ($d_iS_A=0$) and $dS_B=dQ_B/T_B$ ($d_iS_B=0$), respectively \cite{Kondepudi,Groot,Bellac,Yunus}. It should also be noted that $d_iS_{AB}$ is different from $d_iS_A$ and $d_iS_B$ which are the entropies produced inside subsystems $A$ and $B$, respectively. For a closed combined quantum system $AB$ we have
\begin{equation}\label{10a}
dS_{AB}=dS_A+dS_B+dS_C=d_iS_{AB}=0,
\end{equation}
where $dS_C$ is the entropy contribution contained in the \textit{correlations} (such as entanglement) established between the two subsystems. As can be seen from Eq. (\ref{10a}) due to quantum correlations the sum of the entropies of the two subsystems does not equal the entropy of the total system $AB$. Comparing Eq. (\ref{10b}) with Eq. (\ref{10a}) we conclude that the information which was supposed to get lost is actually stored in the quantum correlations established between the two quantum subsystems and this is why no entropy is produced inside the total quantum system $AB$.
If there exist no correlations, $dS_C=0$, we still have $dS_{AB}=dS_A+dS_B=0$ and this means that the unitarity of the dynamics is the main reason for zero entropy production in quantum thermodynamic systems. Staying information inside the total system allows us to reverse the process without leaving any trace on the surroundings. By leaving a trace on the surroundings we mean changing the entropy of the surroundings. A process that can be reversed without leaving any trace on the surroundings is called reversible \cite{Kondepudi,Groot,Bellac,Yunus}. Therefore a process is reversible if and only if $d_iS=0$. In classical thermodynamics sometimes there is no way for a process to be reversed without leaving any trace on the surroundings. These processes are called irreversible processes for which $d_iS\neq 0$. The free expansion of an isolated gas is a bright example of this case which cannot be reversed to its initial state without changing the entropy of the surroundings. Now that we have shown no quantum system is capable of producing any entropy in the interior of itself the generalized temperature of a thermodynamic non-equilibrium quantum system can be defined.
\begin{thm}\label{Theorem2}
The generalized temperature $T(t)$ of a non-equilibrium quantum thermodynamic system connects the change in the entropy of the quantum system $dS(\rho)$ to the exchanged heat $dQ$ with its surroundings as
\begin{equation}\label{12}
dS(\rho)=\dfrac{dQ}{T(t)}.
\end{equation}
\end{thm}
This is plausible because in classical thermodynamics whenever the entropy of a system can be defined the system has also a well-defined temperature and vice versa \cite{Kondepudi,Groot,Bellac,Yunus}. Hence Theorem \ref{Theorem2} states that since the entropy of a quantum system always exists, whether the system is in an equilibrium state or not, a well-defined temperature can be defined. Using Eq. (\ref{10a}) and Theorem \ref{Theorem2} we get
\begin{equation}\label{12a}
dS_C(\rho)=-(\dfrac{dQ_A}{T_A(t)}+\dfrac{dQ_B}{T_B(t)}).
\end{equation}
One may assert that if the entropy production of a quantum system is always zero all the quantum heat engines should have Carnot efficiency. This is not the case because for a non-equilibrium system the generalized temperature is generally time-dependent and the efficiency of the engine becomes
\begin{equation}\label{13}
\eta=1-\dfrac{\Delta Q_c}{\Delta Q_h}=1-\dfrac{\int T_c(t)dS_c(\rho)}{\int T_h(t)dS_h(\rho)},
\end{equation}
where $T_{c(h)}$ and $S_{c(h)}(\rho)$ are the generalized temperature and the entropy of the system during the interaction with the cold (hot) reservoir, respectively. Only when the system is in thermal equilibrium the efficiency becomes Carnot efficiency.
\newline
\newline
\textit{Summary}. In this work we have shown that fundamental differences appear in quantum thermodynamics compared to the classical thermodynamics. We observed that unlike classical thermodynamics no entropy can be produced in the interior of a quantum system. We saw that the information which is supposed to get lost is in fact stored in the quantum correlations between subsystems. At last zero entropy production led to the idea of defining the generalized temperature for a non-equilibrium quantum system.

\section{APPENDIX I}\label{AppendixA}
Von Neumann designed an experiment \cite{Neumann,Goold} which accounts for the work cost of erasing the state of a gas of $n$ atoms, initially in a state $\rho^{\otimes n}$ with $\rho=\sum_{i}p_i|\psi_i\rangle\langle\psi_i|$ by transforming it into a pure state $|\psi_i\rangle^{\otimes n}$ using a reversible process.
\begin{figure}[h]
\centering
\includegraphics[width=5cm]{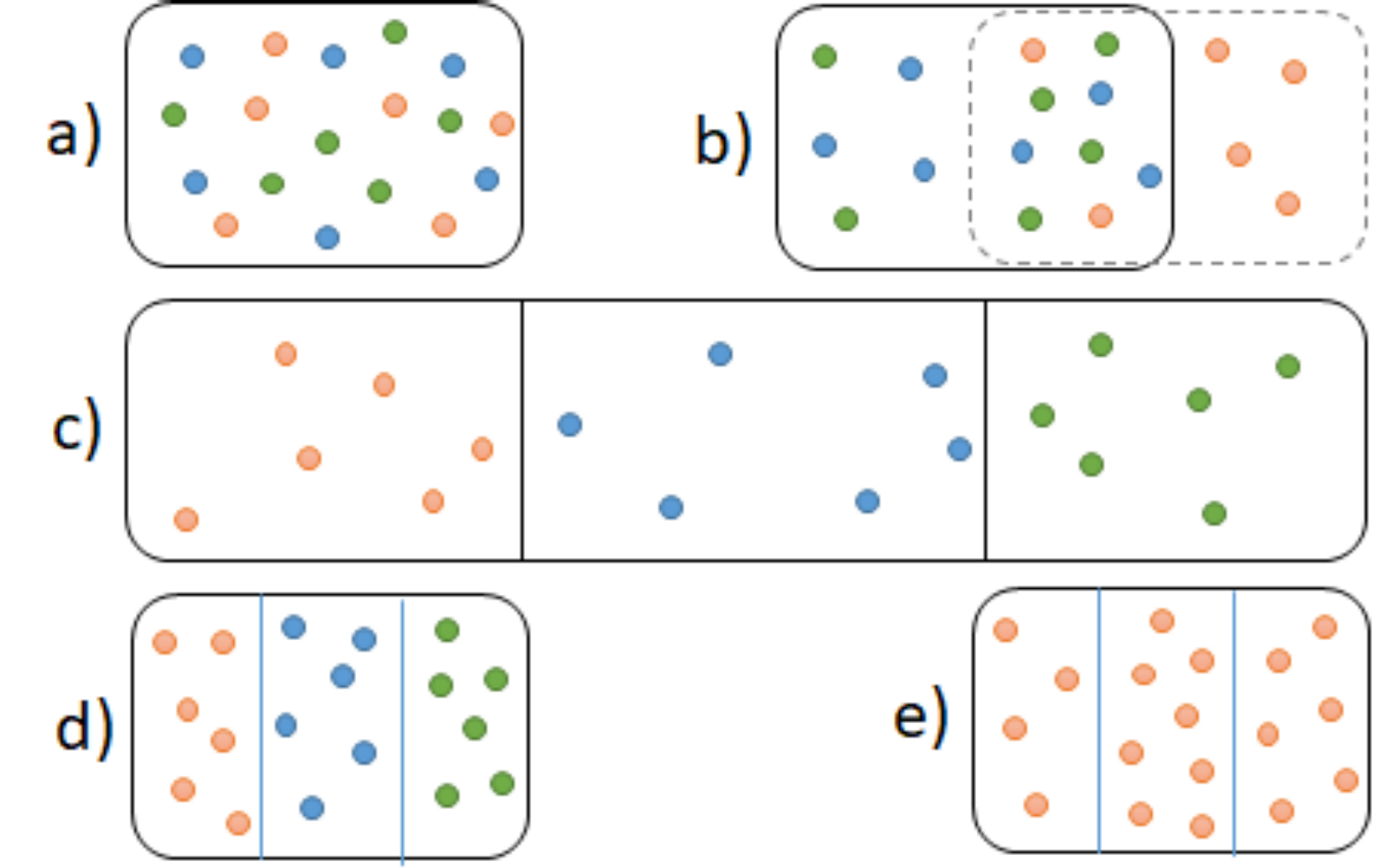}
\caption{(Color online) The steps during which the state of a gas of $n$ atoms, initially in a state $\rho^{\otimes n}$ with $\rho=\sum_{i}p_i|\psi_i\rangle\langle\psi_i|$ is erased by transforming it into a pure state $|\psi_i\rangle^{\otimes n}$ by means of a reversible process.}
\label{fig:Fig3.eps}
\end{figure}
\newline
It consists of 3 steps (as depicted in Fig. (\ref{fig:Fig3.eps})): 1. \textit{Separation of the species}: the atoms in different states $|\psi_1\rangle,...,|\psi_m\rangle$ inside a box of volume $V$ are separated in different boxes of the same volume $V$ by means of semi-permeable walls (from a to b and finally c). Notice that no work has been done and no heat has been exchanged. 2. \textit{Compression}: every gas $|\psi_i\rangle$ is isothermally compressed to a volume $V_i=p_iV$ (from c to d). The mechanical work done in that process is $W_i=np_i\ln(V_i/V)=p_i\ln p_i$. The total entropy increase per particle of the process is $\Delta S=\sum_{i}p_i\ln p_i$. 3. \textit{Unitary transformation}: every gas is put in the $|\psi_1\rangle$ state by applying different unitary transformation $|\psi_i\rangle\rightarrow|\psi_1\rangle$ which are taken for free (from d to e). As the entropy of the final state is zero, the entropy of the initial state reads
\begin{equation}\label{app}
S(\rho)=-tr\{\rho\ln\rho\}.
\end{equation}
\end{document}